\documentclass[fancyhdr]{article}

\usepackage{authblk}
\usepackage[utf8]{inputenc}
\usepackage[T1]{fontenc}
\usepackage{pslatex}
\usepackage[english]{babel}
\usepackage{fancyhdr}
\usepackage{hyperref}
\usepackage{amsmath,amssymb,amsfonts,amsthm}
\usepackage{graphicx}
\usepackage{amsmath}
\usepackage{amsthm}
\usepackage{amsfonts}
\usepackage{mathtools}
\usepackage{amssymb}
\usepackage{algorithm}
\usepackage{algorithmic}
\usepackage{microtype}
\usepackage{rotating}
\usepackage{todonotes}
\usepackage{tikz}
\usepackage{tikz-qtree}
\usepackage{xparse}
\usepackage{appendix}
\usepackage{url}
\usepackage{tabularx}
\usepackage{paralist}
\usepackage{booktabs}
\usepackage{multirow}
\usepackage{subcaption}
\usepackage{cancel}
\usetikzlibrary{calc,shapes.callouts,shapes.arrows}
\usetikzlibrary{automata}
\usetikzlibrary{positioning}
\usetikzlibrary{decorations.pathreplacing}
\usetikzlibrary{decorations.pathmorphing}
\usetikzlibrary{arrows}
\usetikzlibrary{shapes}

\newcommand{\Z}{\mathbb{Z}}
\newcommand{\F}{\mathbb{F}}

\newtheorem*{problem*}{Problem}

\providecommand{\keywords}[1]{\textbf{\textit{Keywords }} #1}

\begin{document}

\title{IDEM Enough? Evolving Highly Nonlinear Idempotent Boolean Functions}

\author[1,2]{Claude Carlet}
\author[3]{Marko \DH urasevic}
\author[3]{Domagoj Jakobovic}
\author[4]{Luca Mariot}
\author[3,5]{Stjepan Picek}

\affil[1 ]{{\normalsize University of Paris 8, Saint-Denis, France}}

\affil[2 ]{{\normalsize University of Bergen, Bergen, Norway}

    {\small \texttt{claude.carlet@gmail.com}}}

\affil[3 ]{{\normalsize Faculty of Electrical Engineering and Computing, University of Zagreb, Unska 3, Zagreb, Croatia} \\

{\small \texttt{marko.durasevic@fer.hr, domagoj.jakobovic@fer.hr}}}

\affil[4 ]{{\normalsize Semantics, Cybersecurity and Services Group, University of Twente, 7522 NB Enschede, The Netherlands} \\
	
	{\small \texttt{l.mariot@utwente.nl}}}

\affil[5 ]{{\normalsize Digital Security Group, Radboud University, Postbus 9010, 6500 GL Nijmegen, The Netherlands} \\
	
	{\small \texttt{stjepan@computer.org}}}
	
\maketitle

\begin{abstract}
Idempotent Boolean functions form a highly structured subclass of Boolean functions that is closely related to rotation symmetry under a normal-basis
representation and to invariance under a fixed linear map in a polynomial basis.
These functions are attractive as candidates for cryptographic design, yet their additional algebraic constraints make the search for high nonlinearity substantially more difficult than in the unconstrained case.
In this work, we investigate evolutionary methods for constructing highly nonlinear idempotent Boolean functions for
dimensions $n=5$ up to $n=12$ using a polynomial basis representation with canonical primitive polynomials. 
Our results show that the problem of evolving idempotent functions is difficult due to the disruptive nature of crossover and mutation operators. Next, we show that idempotence can be enforced by encoding the truth table on orbits, yielding a compact genome of size equal to the number of distinct squaring orbits. 
\end{abstract}


\section{Introduction}
\label{sec:intro}
Boolean functions with strong spectral and cryptographic properties are central building blocks in symmetric cryptography, including stream ciphers and block ciphers~\cite{carlet_2021}. They also play a key role in sequence design~\cite{1056589}, coding theory~\cite{KERDOCK1972182,MacWilliams-Sloane}, and combinatorics~\cite{Rothaus}.
For $n$ input variables, the set of Boolean functions has cardinality $2^{2^n}$, which makes exhaustive search infeasible already for moderate $n$ (e.g., $n=7$). Consequently, the design of Boolean functions often relies on (i) provable constructions and/or (ii) heuristic and evolutionary search guided by
cryptographic criteria.
In cryptographic applications, ``good'' usually means simultaneously satisfying several constraints, such as high nonlinearity (Walsh-Hadamard spectral flatness), desirable autocorrelation, sufficient algebraic degree, and often balancedness or
near-balancedness depending on the use case~\cite{carlet_2021}. Since only a small fraction of all Boolean functions fulfill such requirements, structured subclasses and systematic construction principles are of extreme importance.

A classical way to impose structure is to require invariance under algebraic symmetries. In this work, we focus on idempotent Boolean functions, defined in the univariate representation over $\mathbb{F}_{2^n}$ by the Frobenius invariance 
  $f(x)=f(x^2), \forall x\in\mathbb{F}_{2^n}.$
This condition implies that the truth table is constant on the orbits of the Frobenius map $x\mapsto x^2$. Hence, idempotence reduces the effective degrees of freedom: instead of choosing $2^n$
independent output bits, an idempotent function is determined by one bit per Frobenius orbit. This observation is particularly beneficial for evolutionary search, as it allows encodings that guarantee feasibility by construction and avoid disruptive repair operators.

While the general body of work on evolving Boolean functions is rich (see Section~\ref{sec:related}), evolutionary studies that explicitly incorporate algebraic symmetries are comparatively rare. Most notably, rotation-symmetric Boolean functions have received attention, largely because cyclic rotations are easy to express and exploit. There is a well-known equivalence between rotation symmetry and idempotence when a normal basis is used to represent $\mathbb{F}_{2^n}$. In that coordinate system, Frobenius squaring corresponds to a cyclic shift of coordinates (see
Section~\ref{sec:background}). However, in a polynomial basis - as obtained from a fixed irreducible (or primitive) polynomial - the Frobenius map corresponds to a generally nontrivial linear transformation. As a result, the orbit membership (and thus the induced constraints on truth table positions) differs from the rotation case. Working in a polynomial basis is attractive for reproducibility and
implementability: it provides a canonical representation (via a chosen polynomial) and aligns with common finite-field implementations.

In this paper, we investigate the evolutionary construction of idempotent highly nonlinear Boolean functions defined through a polynomial basis. To the best of our knowledge, this setting has not been studied in prior evolutionary work. We consider three encodings and two fitness functions across eight input sizes, and we demonstrate that a naive (unrestricted) truth table encoding performs poorly due to the disruptive effect of crossover and mutation under the idempotence constraint. In contrast, an orbit-based restricted encoding---which evolves only one bit per Frobenius orbit and expands it to a full truth table---makes the problem tractable and consistently yields idempotent functions that are highly nonlinear for all tested sizes.

Our main contributions are as follows:
\begin{itemize}
  \item We analyze the difficulty of evolving highly nonlinear idempotent Boolean functions and identify mechanisms that hinder standard   evolutionary operators, providing an explanation of why naive encodings tend to stagnate.
  \item We propose and evaluate a restricted, orbit-based encoding that enforces idempotence by construction, enabling effective search for highly nonlinear functions while preserving the ability to apply standard evolutionary operators.
  \item We provide a systematic empirical study across multiple dimensions, encodings, and fitness functions.
\end{itemize}

The rest of this paper is organized as follows.
Section~\ref{sec:background} provides information about Boolean functions and their representations, as well as idempotent functions. Section~\ref{sec:related} provides a brief overview of related works. Section~\ref{sec:setup} details our experimental setup. Section~\ref{sec:results} provides experimental results and a discussion.
Finally, Section~\ref{sec:conclusions} summarizes the main contributions of the paper and suggests possible future work directions.

\section{Background}
\label{sec:background}

We denote by $\mathbb F_2=\{0,1\}$ the finite field with two elements, equipped with XOR (sum) and logical AND (multiplication), respectively. The $n$-dimensional vector space over $\mathbb F_2$ is denoted by $\mathbb F_2^n$, consisting of all $2^n$ binary vectors of length $n$. Given $a, b \in \mathbb F_2^n$, their inner product equals $a\cdot b = \bigoplus_{i=1}^{n} a_{i}b_{i}$ in $\mathbb F_{2}^n$. A Boolean function of $n$ variables is a mapping $f: \F_2^n \to \F_2$. 

\subsection{Boolean Functions - Representations and Properties}

\paragraph{Truth Table Representation.}
A Boolean function $f: \F_2^n \to \F_2$ can be uniquely represented by using its truth table. The truth table (TT) of a Boolean function $f$ is the list of pairs $(x, f(x))$ of input vectors $x \in \F_2^n$ and function outputs $f(x) \in \F_2$. Once a total order has been fixed on the input vectors of $\F_2^n$ (most commonly, the lexicographic order), the truth table can be identified only by the $2^n$-bit function output vector.

\paragraph{Walsh-Hadamard Transform.}
The Walsh-Hadamard transform $W_{f}: \F_2^n \to \Z$ is another unique representation of a Boolean function $f$. The Walsh-Hadamard transform measures the correlation between $f$ and the linear functions $a\cdot x$, for all $a \in \mathbb F_2^n$:
\begin{equation}
W_{f} (a) = \sum\limits_{x \in \mathbb{F}_{2}^{n}} (-1)^{f(x) \oplus a\cdot x},
\end{equation}
with the sum calculated in ${\mathbb Z}$. The Walsh-Hadamard transform is an involution up to a normalization by a constant. As such, one can retrieve the truth table representation of $f$ from the spectrum of its Walsh-Hadamard coefficients $W_f(a)$. The Walsh-Hadamard transform allows easy characterization of several cryptographic properties. 

\paragraph{Nonlinearity}
The minimum Hamming distance between a Boolean function $f$ and all affine functions is the nonlinearity of $f$, which is calculated from the Walsh-Hadamard spectrum as follows~\cite{carlet_2021}:
\begin{equation}
\label{eq:nonlinearity}
nl_{f} = 2^{n - 1} - \frac{1}{2}\max_{a \in \mathbb{F}_{2}^{n}} \left \{ |W_{f}(a)| \right \}.
\end{equation}

Every $n$-variable Boolean function $f$ satisfies the covering radius bound:
\begin{equation}
\label{eq_boolean_covering}
    nl_{f} \leq 2^{n-1}-2^{\frac n 2 - 1}.
\end{equation}
Eq.~\eqref{eq_boolean_covering} cannot be tight when $n$ is odd, which means that the maximal nonlinearity is possible for even $n$ only. Boolean functions in an even number of variables that achieve maximal nonlinearity are called bent Boolean functions. 
On the other hand, the maximal possible nonlinearity for odd-sized Boolean functions equals~\cite{carlet_2021}:
\begin{equation}
    nl_f \leq 2^{n-1}-2^{\frac {n-1}{2}}.
\end{equation}

The maximal possible nonlinearity for  Boolean functions in odd sizes lies between~\cite{carlet_2021}:
$2^{n-1}-2^{\frac{n-1}{2}}$ and $2\lfloor2^{n-2}-2^{\frac n 2-2}\rfloor$. The lower bound is called the quadratic bound, and the highest nonlinearity for odd-sized Boolean functions is strictly larger than the quadratic bound for $n>7$~\cite{carlet_2021}.

\subsection{Polynomial Basis}
Let us fix a primitive polynomial\footnote{A primitive polynomial is an irreducible polynomial whose root generates the entire multiplicative group of the extension field.} $p_n(x)\in\mathbb{F}_2[x]$ of degree $n$ and let $\alpha$ be a root of $p_n(x)$.
We work with the extension field $\mathbb{F}_{2^n}$, realized as the quotient $\mathbb{F}_2[x]/(p_n(x))$. 

In the \emph{polynomial basis} $(1,\alpha,\alpha^2,\dots,\alpha^{n-1})$, any $x\in\mathbb{F}_{2^n}$ is represented as
\[
x=\sum_{i=0}^{n-1} x_i\alpha^i,\qquad x_i\in\mathbb{F}_2.
\]

Let us write a monic degree-$n$ polynomial as
$p(x)=x^n + a_{n-1}x^{n-1}+\cdots+a_1x+a_0$ with $a_i\in\mathbb{F}_2$ and $a_0=1$.
We order candidates by the coefficient vector $(a_{n-1},a_{n-2},\dots,a_1,a_0)$ in lexicographic order. For each $n$, we select the first polynomial in this order that is primitive.

Using this polynomial basis representation, a Boolean function can also be defined as a mapping $f: \mathbb{F}_{2^n} \to \mathbb{F}_2$ (i.e., mapping elements of the extension field $\mathbb{F}_{2^n}$ to $\mathbb{F}_2$, instead of vectors in $\mathbb{F}_2^n$). In what follows, we refer to $F: \mathbb{F}_{2}^n \to \mathbb{F}_2$ as the \emph{multivariate} representation of the function, while $f: \mathbb{F}_{2^n} \to \mathbb{F}_2$ is its \emph{univariate representation}.

\subsection{Normal Basis}
A normal basis is of the form $(\beta,\beta^2,\beta^{2^2},\dots,\beta^{2^{n-1}})$ for a normal element\footnote{A normal element is an element of a finite-field extension whose Frobenius conjugates form a basis.} $\beta$. In a normal basis, the Frobenius map $x\mapsto x^2$ corresponds to a cyclic rotation of coordinates. In the polynomial basis, it corresponds to a general $\mathbb{F}_2$-linear map.

\subsection{Idempotent Boolean Functions}

A Boolean function $f:\mathbb{F}_{2^n}\to\mathbb{F}_2$ is idempotent if
\[
f(x)=f(x^2)\qquad\forall x\in\mathbb{F}_{2^n}.
\]
Equivalently, $f$ is constant on the Frobenius orbits\footnote{A Frobenius orbit is the set obtained by repeatedly applying the Frobenius automorphism to an element of a finite field.}, defined as $\mathcal{O}(x)=\{x,x^2,x^{2^2},\dots\}$.
This yields an immediate parametrization: assign one output bit per orbit, then extend constantly over it.

A Boolean function over $\mathbb{F}_2^n$ is called rotation symmetric if it is invariant under any cyclic shift of the input coordinates, that is if $f(x_0, x_1, \ldots,  x_{n-1}) = f(x_{n-1},  x_0, x_1, \ldots, x_{n-2})$ for all vectors $(x_0, x_1, \ldots, x_{n-1}) \in \F_2^n$.

The link between idempotence and rotation symmetry becomes clear once a normal basis is used to identify $\mathbb{F}_{2^n}$ with $\mathbb{F}_2^n$. If $x$ is represented as $x=\sum_{j=0}^{n-1} x_j \beta^{2^j}$ in a normal basis $(\beta,\beta^2,\ldots,\beta^{2^{n-1}})$, then squaring acts as a cyclic shift of coordinates: $x^2=\sum_{j=0}^{n-1} x_j \beta^{2^{j+1}}$, which corresponds to rotating $(x_0,x_1,\ldots,x_{n-1})$ by one position (up to direction conventions). Consequently, the idempotence condition $f(x)=f(x^2)$ translates into invariance under cyclic coordinate shifts in the associated multivariate representation $F(x_0,\ldots,x_{n-1})$, i.e., rotation symmetry. 

In this sense, idempotent functions over $\mathbb{F}_{2^n}$ and rotation symmetric Boolean functions over $\mathbb{F}_2^n$ describe the same class once a normal basis is fixed, while in other bases (e.g., polynomial bases), the same Frobenius invariance is expressed as invariance under a different, generally nontrivial linear coordinate transformation.

\subsubsection{Checking Idempotence in Polynomial Basis via a Linear Map}
In polynomial basis, squaring is $\mathbb{F}_2$-linear:
\[
\left(\sum_{i=0}^{n-1}x_i\alpha^i\right)^2 = \sum_{i=0}^{n-1} x_i\alpha^{2i}\ \bmod\ p_n(\alpha).
\]
Thus, there exists a linear map $S_n:\mathbb{F}_2^n\to\mathbb{F}_2^n$ (or equivalently, an $n\times n$ binary matrix) such that
\[
\mathrm{coords}(x^2) = S_n\,\mathrm{coords}(x).
\]

With the word $X=x_{n-1}\cdots x_0$ and $Y=S_n(X)$ (both in the same bit ordering), a multivariate Boolean function
$F:\mathbb{F}_2^n\to\mathbb{F}_2$ representing $f$ is idempotent if and only if:
\begin{equation}\label{eq:idempotence-check}
F(X) = F(S_n(X))\qquad \forall X\in\mathbb{F}_2^n.
\end{equation}
For truth tables, Eq.~\eqref{eq:idempotence-check} is checked by iterating over all inputs and comparing the output with the output at the mapped input. Equivalently, we can enumerate the cycles of $S_n$ and verify that $F$ is constant on each cycle.

\subsection{Why Polynomial Basis?}
A natural choice for studying idempotent Boolean functions is a normal basis, because in a normal basis, the Frobenius map $x\mapsto x^2$ corresponds to a cyclic rotation of coordinates. Consequently, the idempotence constraint $f(x)=f(x^2)$ becomes rotation symmetry in the multivariate representation under a normal basis.
In this paper, however, we adopt a polynomial basis representation of $\mathbb{F}_{2^n}$. Next, we list several reasons for this choice.

\begin{enumerate}
    \item Polynomial bases are the default representation in many software and hardware implementations of $\mathbb{F}_{2^n}$ arithmetic: elements are stored as $n$-bit words (polynomial coefficients), and multiplication is implemented via carryless multiplication followed by reduction modulo a fixed irreducible (often primitive) polynomial.
This makes the polynomial-basis model a practical target for designing and testing functions intended for deployment.
Moreover, by fixing for each $n$ a canonical primitive polynomial (in our case, the lexicographically first one), we remove a common ambiguity: in a normal-basis approach, one must also specify the normal element $\beta$ (and many choices exist), whereas in a polynomial basis a published choice of $p_n(x)$ completely determines the coordinate system.
As a result, our idempotence checks and reported functions can be reproduced without reconstructing a normal basis or validating that the same $\beta$ was used.

\item Although a normal basis makes squaring cheap (a coordinate rotation), a polynomial basis does not make the problem intractable. Indeed, the Frobenius map remains $\mathbb{F}_2$-linear in \emph{any} basis.
Hence, for each $n$, there exists a fixed linear map (binary matrix) $S_n$ such that
$\mathrm{coords}(x^2)=S_n\,\mathrm{coords}(x)$.
Once $S_n$ is derived (or precomputed) for the chosen $p_n(x)$, enforcing or verifying idempotence reduces to evaluating
\[
F(X)=F(S_n(X))\qquad \forall X\in\mathbb{F}_2^n,
\]
which is computationally inexpensive (in practice, a small XOR network per evaluation, or constant-on-orbits constraints in a truth-table representation).
As such, we exchange a shift operation for a fixed linear map, while keeping the overall cost small and easily reproducible.

\item Even though idempotence testing only requires squaring, many constructions and secondary criteria in Boolean function design are most naturally expressed using finite field operations such as multiplication and trace. In a polynomial basis, multiplication is straightforward to implement via modular reduction, without requiring a multiplication table or special structure to achieve acceptable performance.
This simplicity helps keep the experimental pipeline compact and transparent.

\item In a normal basis, idempotence becomes indistinguishable from a specific coordinate symmetry (rotation symmetry), which is a powerful viewpoint but also risks conflating a basis-invariant property (Frobenius invariance) with a basis-dependent
multivariate description (rotation of coordinate positions).
Working in a polynomial basis helps decouple these aspects: idempotence is still exactly Frobenius invariance, but it is expressed through a generally nontrivial linear map $S_n$ rather than a coordinate rotation.
This allows us to study how evolutionary search behaves under the same field-theoretic constraint when the induced symmetry is not ``geometric'' in the coordinate sense, and to distinguish basis-invariant phenomena (idempotence) from basis-dependent ones (the appearance of rotation symmetry).

\item Evolutionary operators define a neighborhood on the search space.
In normal basis, the squaring action is a simple rotation, and the orbit structure can appear highly regular in coordinates. In a polynomial basis, orbits are the same in the field, but their manifestation through $S_n$ can lead to different coordinate mixing behavior, potentially affecting mutation efficacy, diversity maintenance, and the prevalence of local optima. 
Treating the basis choice as an explicit experimental design variable, therefore, provides additional insight into the interaction between algebraic constraints and heuristic optimization.
\end{enumerate}

\paragraph{The Number of Orbits}
Let $S$ be a finite set and let $\theta:S\to S$ be a permutation (bijective map). The orbit of an element $s\in S$ under $\theta$ is the set obtained by iterating $\theta$,
\[
\mathrm{Orb}_\theta(s)=\{s,\theta(s),\theta^2(s),\ldots\}.
\]
Since $S$ is finite, this sequence eventually repeats and $\mathrm{Orb}_\theta(s)$ is a cycle. The orbits form a partition of $S$. For Boolean functions invariant under $\theta$ (i.e., $F(s)=F(\theta(s))$ for all $s$), the function must be constant on each orbit, hence it can be encoded by one bit per orbit.

The number of rotation symmetric and idempotent Boolean functions is smaller than the number of Boolean functions, as the output value remains the same for certain input values. This follows from Burnside's lemma~\cite{burnside}. For a cyclic group action of order $n$ generated by $g$, the number of orbits is
\[
\#\mathrm{Orbits}=\frac{1}{n}\sum_{k=0}^{n-1}\left|\mathrm{Fix}(g^k)\right|,
\qquad
\mathrm{Fix}(g^k)=\{s\in S: g^k(s)=s\}.
\]
For the rotation action, $\rho^k$ fixes exactly those binary vectors with period dividing $\gcd(n,k)$:
\[
\left|\mathrm{Fix}(\rho^k)\right| = 2^{\gcd(n,k)}.
\]
For the Frobenius action, $\varphi^k(x)=x^{2^k}$, and its fixed points in $\mathbb{F}_{2^n}$ form the subfield
$\mathbb{F}_{2^{\gcd(n,k)}}$, whose size is
\[
\left|\mathrm{Fix}(\varphi^k)\right|=\left|\mathbb{F}_{2^{\gcd(n,k)}}\right|=2^{\gcd(n,k)}.
\]
Since the fixed-point counts coincide for every $k$, Burnside's lemma gives the same orbit count in both settings:
\[
\#\mathrm{Orbits}=\frac{1}{n}\sum_{k=0}^{n-1}2^{\gcd(n,k)}.
\]

Therefore, rotation-symmetric Boolean functions and idempotent Boolean functions (for any fixed basis, including a polynomial basis defined by a chosen irreducible/primitive polynomial) have the same number of independent orbit bits, even though the induced orbit partition on truth-table positions may differ.
We provide the number of orbits in Table~\ref{tab:rots}. 
\begin{table}
  \centering
  \caption{The number of orbits for the idempotent Boolean functions. The number of Boolean functions for each dimension then equals $2^{orbits}$.}
  \label{tab:rots}
  \begin{tabular}{cccccccccc}
     &             \\
    \toprule
    variables & 4 &	5 & 6&	7 & 8&	9 & 10& 11 & 12	 \\
    \midrule
    orbits&	6 & 8 & 14&	20 & 36& 60 &	108&	188 & 352	 \\
\bottomrule
  \end{tabular}
\end{table}

\section{Related Work}
\label{sec:related}

Boolean functions can be constructed randomly, by using algebraic constructions, and with (meta)heuristics.
The last direction shows steady development over the years, primarily due to very good results achieved for diverse specific problems, see, e.g.,~\cite{10.1007/978-3-319-16501-1_16, 00190, 10.1145/2908812.2908915, a16110499, 10.1145/3579856.3590337}. 
Note that the problem of designing bent Boolean functions is especially well studied in recent years~\cite{Djurasevic2023}.

Fuller et al. used evolutionary algorithms to design bent Boolean functions, adding random algebraic normal form terms of specific order~\cite{FullerDM03}.
Picek and Jakobovic discussed the computational difficulty of evolving bent Boolean functions in a large number of variables (e.g., more than 12), and they proposed an evolutionary algorithm approach where, instead of bent Boolean functions, one evolves secondary constructions of bent Boolean functions~\cite{10.1145/2908812.2908915}.
On the other hand, Hrbacek and Dvorak explored diverse configurations of Cartesian Genetic Programming to speed up the evolution process and evolve bent Boolean functions up to 16 variables~\cite{10.1007/978-3-319-10762-2_41}.  
Mariot et al. used evolutionary strategies to evolve a secondary construction based on cellular automata for quadratic bent functions~\cite{MariotSLM22}.
Husa and Dobai used linear genetic programming to evolve bent Boolean functions, and they concluded that linear genetic programming copes better with a growing number of function inputs than genetic programming~\cite{10.1145/3067695.3084220}.

To our knowledge, this paper is the first work to examine evolving idempotent functions that are constructed with a polynomial basis. However, there are multiple works considering rotation symmetric Boolean functions as discussed next. 
The first work we are aware of is by Stănică et al., who used simulated annealing to evolve rotation symmetric Boolean functions~\cite{StanicaMC04}. Next, Kavut and Yucel used a steepest-descent-like iterative algorithm to construct imbalanced Boolean functions in $9$ variables, where the authors considered the generalized rotation symmetric Boolean functions~\cite{KAVUT2010341}.
Next, Liu and Youssef experimented with simulated annealing, where the objective was to construct balanced rotation symmetric Boolean functions~\cite{4729749}. 
Wang et al. used genetic algorithms to construct rotation symmetric Boolean functions~\cite{Wang2022}.
Carlet et al. considered evolutionary algorithms to evolve balanced and bent rotation symmetric Boolean functions~\cite{CarletDGJMP24} and argued that the bitstring and floating-point representations work the best.
Carlet et al. used evolutionary algorithms and constructed rotation symmetric (anti)-self-dual bent functions for Boolean functions up to 16 variables~\cite{10.1007/978-3-031-70085-9_27}. 
More recently, Carlet et al. explored evolving highly nonlinear Boolean functions in odd dimensions, where they showed that a genetic algorithm with bitstring representation works best and allows evolving maximally nonlinear functions for certain Boolean function sizes, especially when restricting the search space to rotation symmetric functions~\cite{Carlet2025}.

\section{Experimental Setup}
\label{sec:setup}

\subsection{Representations and Encodings}

\subsubsection{Bitstring Representation (TT)}
The most common way to represent a Boolean function is via its truth table (TT)~\cite{Djurasevic2023}, which is encoded as a bitstring. 
For a Boolean function with $n$ inputs, the truth table is encoded as a bitstring of length $2^n$.
The bitstring represents the Boolean function upon which the algorithm operates directly. Therefore, in this case, the algorithm explores the full space of $n$-variable Boolean functions, which has size $2^{2^n}$. 

\subsubsection{Symbolic Representation (GP)}
The second approach in our experiments uses tree-based genetic programming (GP) to represent a Boolean function in its symbolic form. 
This representation usually achieves the best results when dealing with the evolution of Boolean functions with cryptographic properties~\cite{Djurasevic2023}.
In this case, we represent a candidate solution with a tree whose leaves correspond to the input variables $x_0,\ldots, x_{n-1} \in \F_2$. The tree nodes are Boolean operators that combine the inputs received from their children and forward their output to the respective parent nodes. 
We use the following function set: OR, XOR, AND, IF, and the NOT function that takes a single argument. The function IF takes three arguments and returns the second one if the first one evaluates to true, and the third one otherwise.
This function set is commonly used in previous applications.
The output of the root node is the output value of the Boolean function. The truth table of the function $f: \F_2^n \to \F_2$ is determined by evaluating the tree over all possible $2^n$ assignments of the inputs at the leaves. 

\subsubsection{Unrestricted Encoding.}
In the unrestricted form, the previous representations explore the whole space of Boolean functions, which means they are able to represent a solution that is not idempotent.
In that case, we rely on the fitness function to penalize the invalid solutions, where the penalty term is proportional to the distance to the feasible space.

\subsubsection{Restricted Encoding.}
Since the number of idempotent functions is smaller than the number of all Boolean functions, we can use the orbits in the truth table (sets of entries that must have the same value) to effectively restrict the number of bits needed to represent an idempotent function, and by extension, significantly reduce the problem search space.
We already discussed the number of orbits in Section~\ref{sec:background}. The bitstring length in the restricted encoding will therefore be equal to the number of idempotent orbits. We find all elements of an orbit by repeatedly squaring the input and maintaining the list of visited inputs (where the list is of size $2^n$). Once we have all the representatives, we can use the restricted encoding. 
To extract the complete truth table from the restricted encoding, we simply assign the same truth table value to all orbit members.

Since GP cannot be restricted to evolve only idempotent functions in a straightforward way, the restricted GP encoding is realized in the following way.
The truth table obtained by evaluating the GP tree is not used directly; instead, only a single representative bit is retained for each orbit.
The representative bit is arbitrarily taken to be the one mapped by the smallest input vector in lexicographic order (i.e., the ``leftmost'' one in the truth table representation of the function), and its value is copied over all other truth table entries corresponding to the same orbit.
This way, the idempotence property is enforced; however, the semantic information is partially lost, since multiple different symbolic expressions will be mapped to the same Boolean function.

To summarize, we use two representations (TT/GP) coupled with two encodings (unrestricted/restricted), totaling in four variants.

\subsection{Fitness Functions}
\label{subsec:fit}

In our experiments, we search for highly nonlinear idempotent functions and use two fitness functions to that end.
Both fitness functions include a penalty term, $PEN$, that counts the occurrences in the truth table that do not satisfy the idempotence property.
This constraint is validated by visiting all truth table entries belonging to the same orbit, repeatedly squaring the inputs and checking equality after each operation.
This term is included in the fitness function with a negative sign to act as a penalty in maximization scenarios.
The delta function $\delta_{PEN, 0}$ assumes the value one when $PEN = 0$ and is zero otherwise. 
Only if the penalty term equals zero (indicating a function is idempotent), the nonlinearity value ($nl_f$) is added, and the fitness is maximized:
\begin{equation}
\label{eq:f1}
fitness_1 : -PEN + \delta_{PEN, 0} \cdot (nl_{f}).
\end{equation}

The second fitness function takes the same form, but is designed to consider the whole Walsh-Hadamard spectrum and not only its extreme value (see Eq.~\eqref{eq:nonlinearity}).
In this case, it includes the number of occurrences of the maximal absolute value in the spectrum, denoted as $\#max\_values$.
Since higher nonlinearity corresponds to a \textit{lower} maximal absolute value, we aim for as few occurrences of the maximal value as possible to make it easier for the algorithm to reach the next nonlinearity value.
Therefore, the second fitness function is defined as:
\begin{equation}
\label{eq:bent}
fitness_2 : -PEN + \delta_{PEN, 0} \cdot \left( nl_{f} + \frac{2^n - \#max\_values}{2^n} \right).
\end{equation}
Note that the term $\frac{2^n - \#max\_values}{2^n}$ never reaches the value of $1$ since, in that case, we effectively reach the next nonlinearity level.

\subsection{Algorithms and Parameters}
\label{sec:settings}

\paragraph{Common Experimental Parameters.}
For all representations and encodings, we employ the same evolutionary algorithm: a steady-state selection with a 3-tournament elimination operator. 
In each iteration of the algorithm, three individuals are chosen at random from the population for the tournament, and the worst one in terms of fitness value is eliminated. 
The two remaining individuals in the tournament are used with the crossover operator to generate a new child individual, which then undergoes mutation with individual mutation probability $p_{mut} = 0.5$. The mutated child replaces the eliminated individual in the population.
If not stated otherwise, all experiments use a population size of 500 individuals and the same stopping criterion of $10^6$ evaluations.

\paragraph{Genetic Operators.}
For both bitstring encodings, we use the simple bit mutation and the shuffle mutation. For crossover, we employ one-point and uniform crossover operators. Each time the evolutionary algorithm invokes a crossover or mutation operation, one of these operators is randomly selected.

For the symbolic (GP) representation, the genetic operators used in our experiments with tree-based GP are simple tree crossover, uniform crossover, size fair, one-point, and context preserving crossover~\cite{poli08:fieldguide} (selected at random), and subtree mutation.
Applying a different randomly selected genetic operator in each mutation and crossover event is based on preliminary results indicating better convergence when using a diverse set of operators.

\paragraph{Local Search.}
Additionally, both representations can be used with a generic local search operator that works as follows: the operator acts on a single solution and performs a predefined number of mutations. 
If a better solution is found, the new solution immediately replaces the original one, and the operator is applied again.
If no better solution is found after the given number of mutations, the operator terminates.
The operator is applied after each generation and acts upon the current best solution and a number of random solutions.
In our experiments, the number of solutions undergoing local search was set to 1\% of the population size, and the number of trials (random mutations per individual) was set to 25.

\section{Experimental Results}
\label{sec:results}

In Table~\ref{tab:bestnl} we show only the best obtained fitness values for all function sizes and all approaches; for $fitness_2$ we truncate the non-integer part, which is always smaller than 1.
Therefore, if the value is positive, the integer part of the fitness is equal to the obtained nonlinearity value.
The notation denotes the representation (TT, GP), whether the encoding is further restricted (TT\_R, GP\_R), and which fitness function was used ($fit_1$, $fit_2$).
The bottom row shows the optimal or best known nonlinearity values; in the case of an even number of variables, these values correspond to bent functions (and as such, best-possible nonlinearity values).

\begin{figure}
    \centering
    \includegraphics[trim=3.5cm 4cm 0cm 0cm, width=1\linewidth]{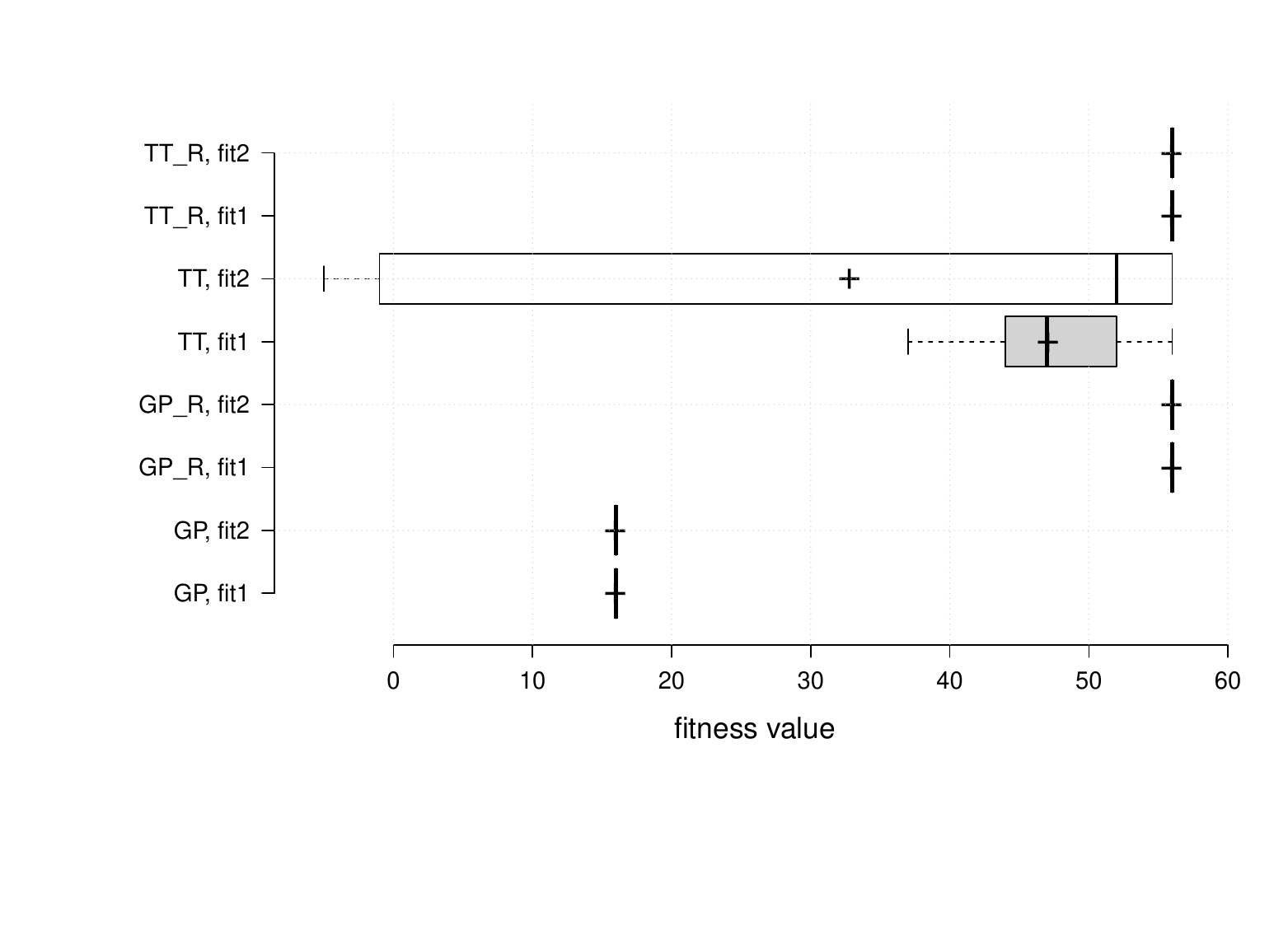}
    \caption{Comparison for $n = 7$ variables}
    \label{fig:n=7}
\end{figure}

\begin{figure}
    \centering
    \includegraphics[trim=3.5cm 4cm 0cm 0cm, width=1\linewidth]{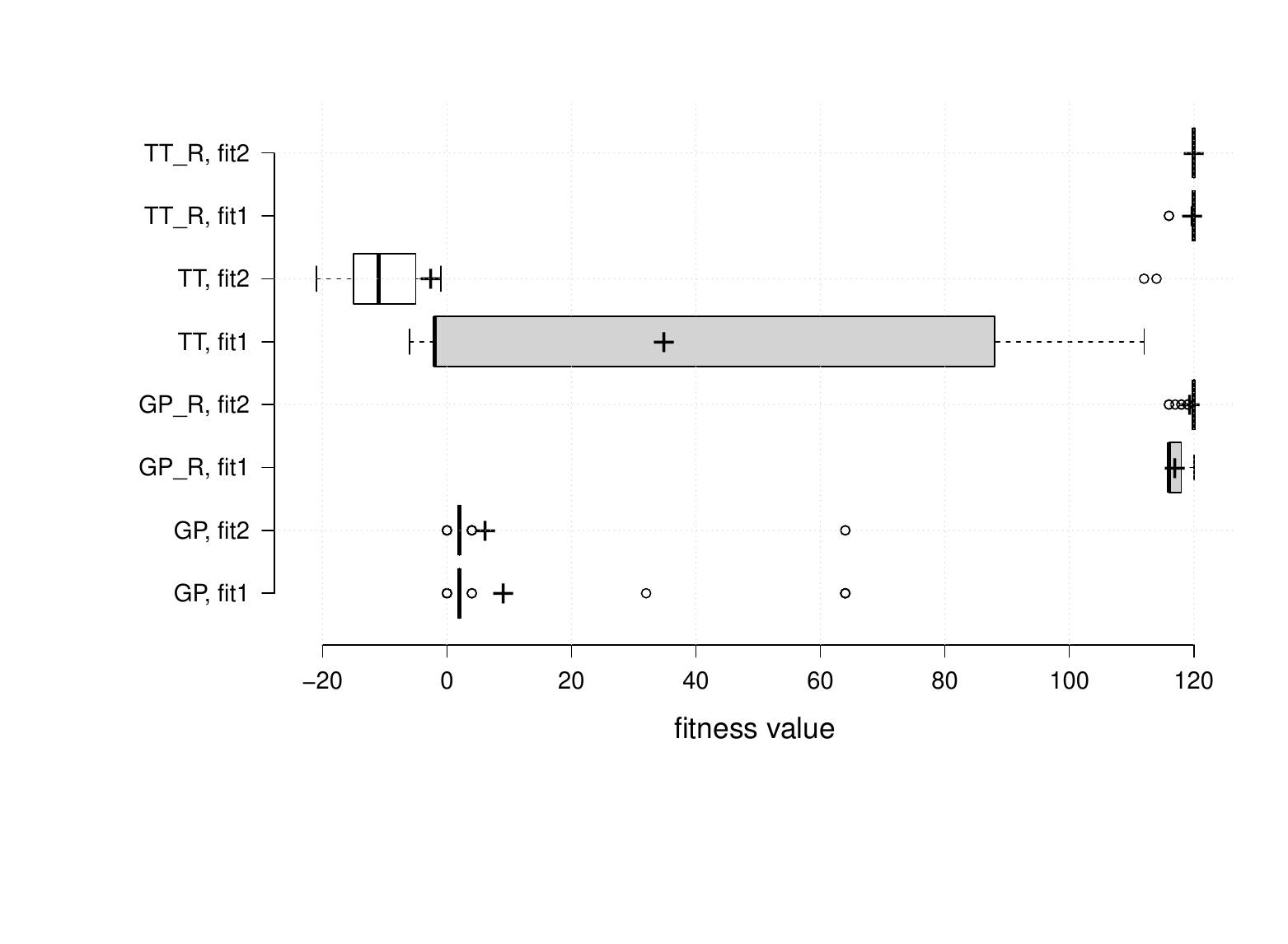}
    \caption{Comparison for $n = 8$ variables}
    \label{fig:n=8}
\end{figure}

\begin{figure}
    \centering
    \includegraphics[trim=3.5cm 4cm 0cm 0cm, width=1\linewidth]{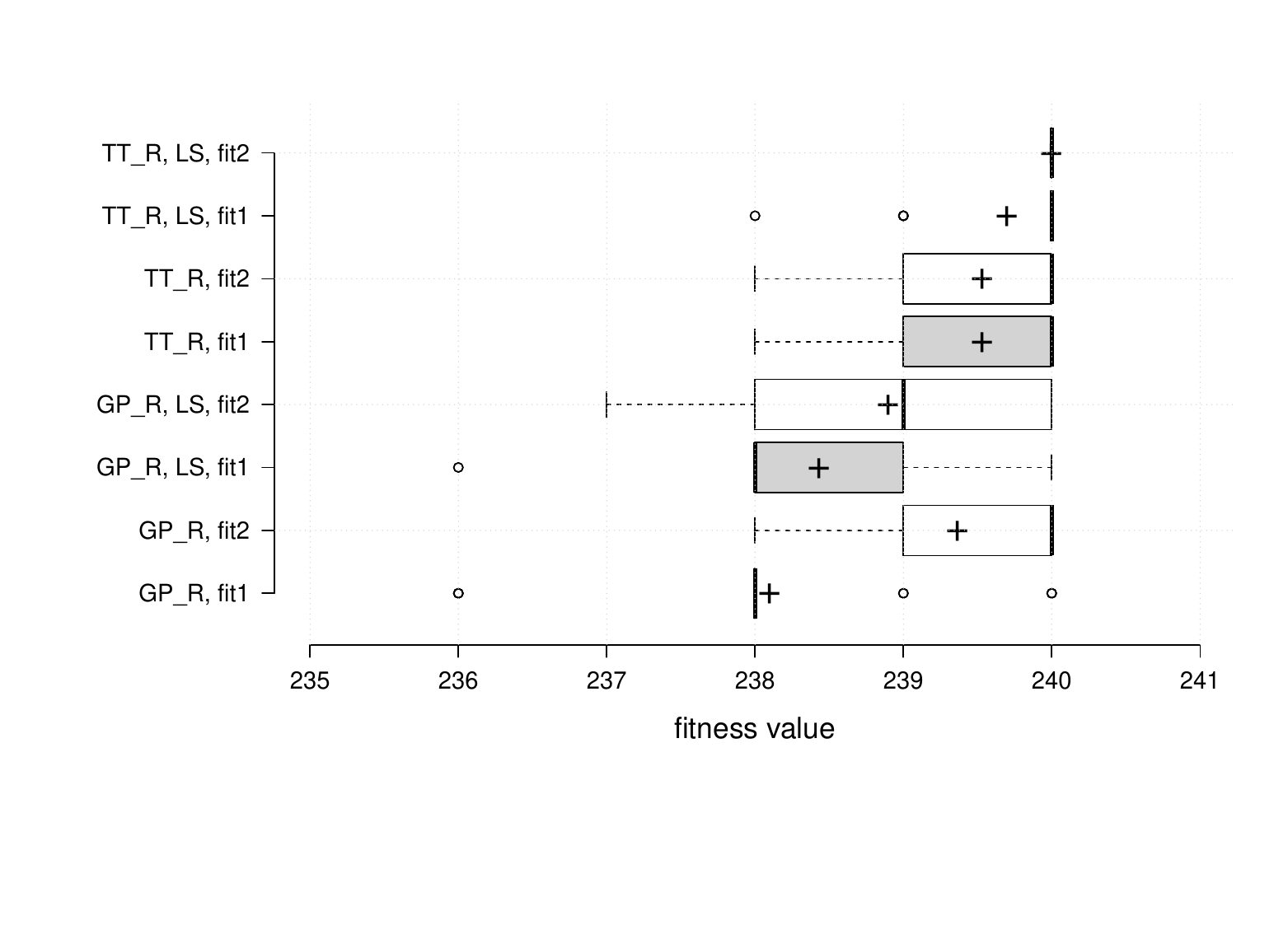}
    \caption{Comparison for $n = 9$ variables}
    \label{fig:n=9}
\end{figure}

\begin{figure}
    \centering
    \includegraphics[trim=3.5cm 4cm 0cm 0cm, width=1\linewidth]{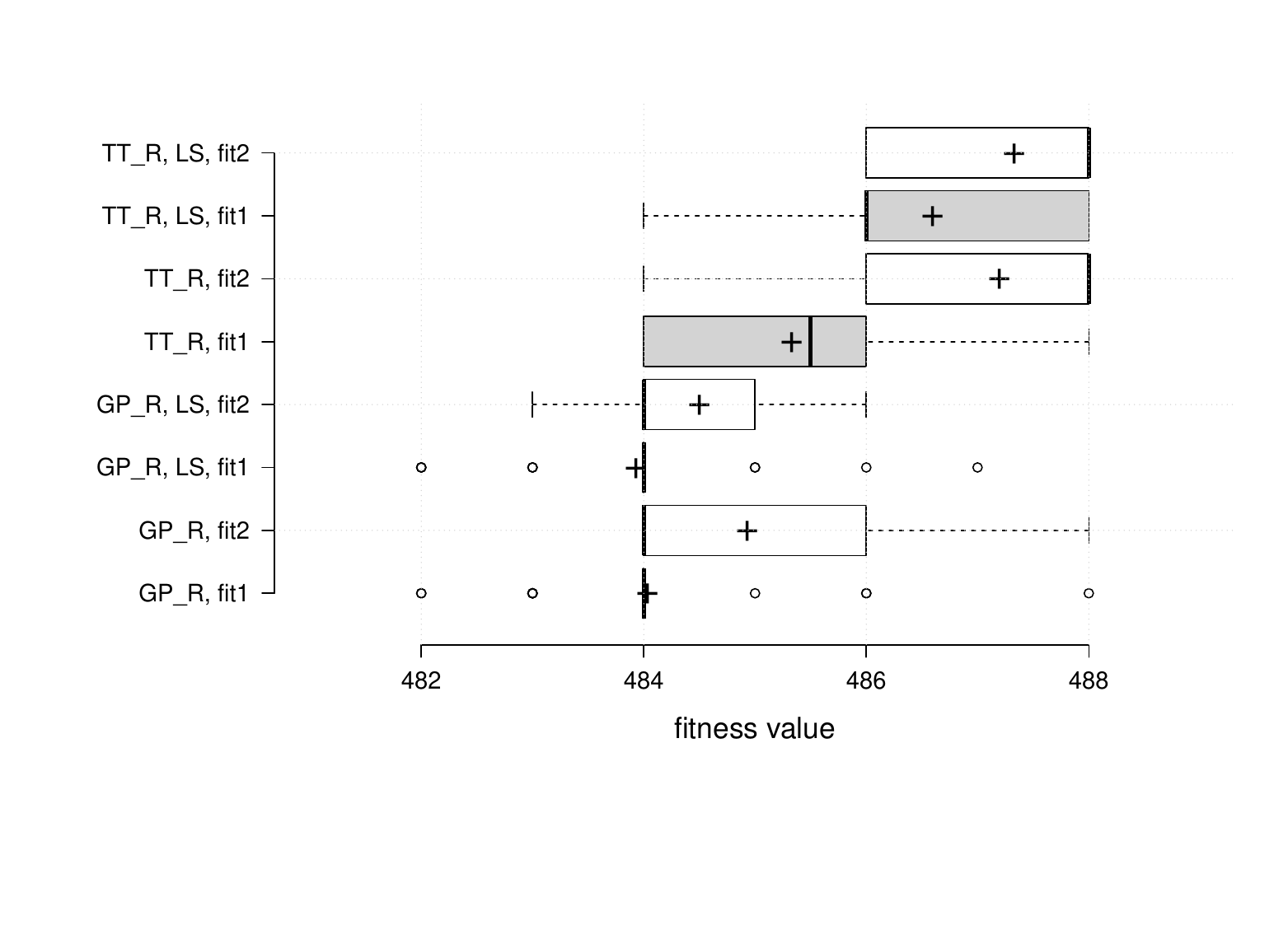}
    \caption{Comparison for $n = 10$ variables}
    \label{fig:n=10}
\end{figure}

\begin{figure}
    \centering
    \includegraphics[trim=3.5cm 4cm 0cm 0cm, width=1\linewidth]{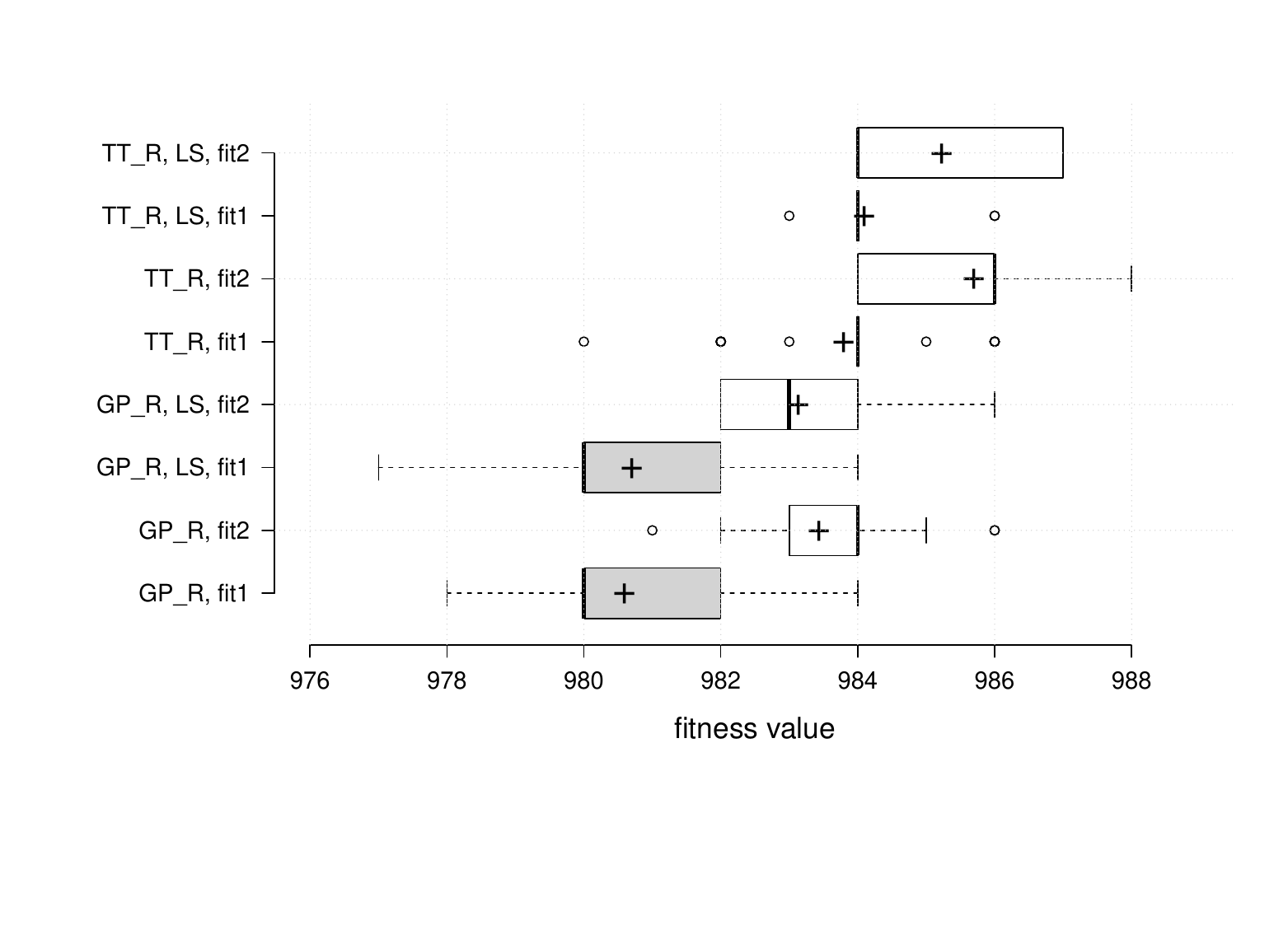}
    \caption{Comparison for $n = 11$ variables}
    \label{fig:n=11}
\end{figure}

\begin{figure}
    \centering
    \includegraphics[trim=3.5cm 4cm 0cm 0cm, width=1\linewidth]{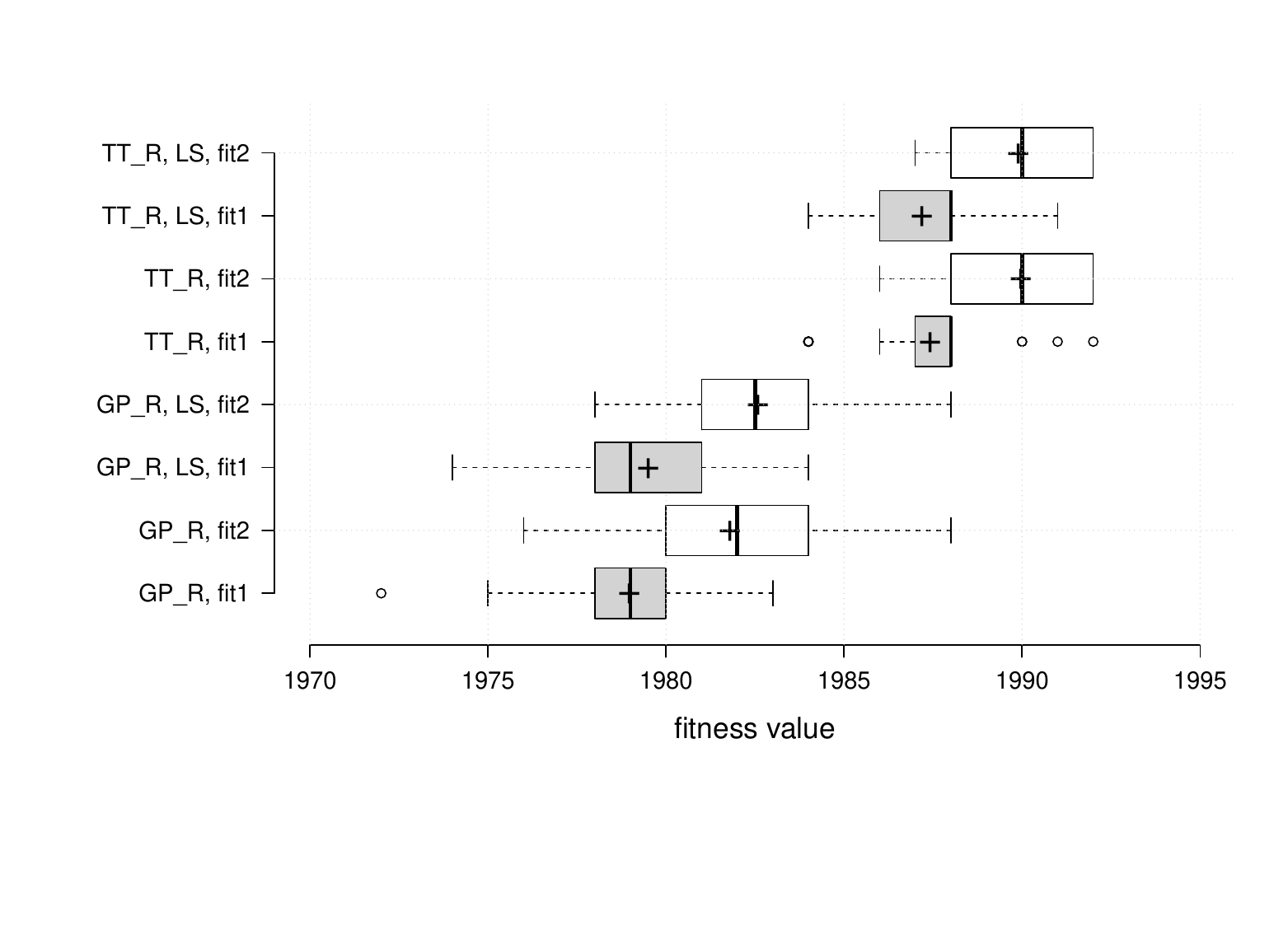}
    \caption{Comparison for $n = 12$ variables}
    \label{fig:n=12}
\end{figure}

Regarding the encoding methods, it is immediately apparent that the unrestricted encoding is significantly worse than the restricted one; this is visible from Figures~\ref{fig:n=7} and~\ref{fig:n=8}, which show boxplot results for all encodings in 7 and 8 variables.
The restricted encodings succeed in finding optimal (bent) nonlinearity in every combination, while the unrestricted variants often fail in constructing an idempotent function in the first place (seen as negative fitness values), and already for 8 variables never reach the optimal nonlinearity values.
Because of this, we omit the unrestricted variants in the figures for larger numbers of variables.
For the same reason, we apply the local search only to the restricted variants; the results for a larger number of variables are shown in Figures~\ref{fig:n=9},~\ref{fig:n=10},~\ref{fig:n=11}, and~\ref{fig:n=12}.

To evaluate the existence of statistically significant differences between the tested encodings, fitness functions, and the use of the local search (LS) operator, the Mann--Whitney U test is applied with a significance level of 0.05. The analysis is restricted to problem size 12, which represents the largest instance considered, as the box plots show consistent behavior across the other large problem sizes.
The comparison of solution representations clearly indicates that the TT representation achieves significantly superior results, with a p-value equal to 0. Likewise, the results obtained using the fitness function $fit_2$ are statistically significantly better than those obtained with $fit_1$, yielding a p-value of $2 \times 10^{-6}$, regardless of the representation.
Finally, the comparison between algorithms employing the LS operator and those without it produces a p-value of 0.7328, revealing no statistically significant difference. Therefore, the application of the LS operator does not provide any improvement in the results.

The fact that bitstring representation achieves better results than genetic programming might be attributed to the nature of restricted encoding in GP.
In the presented GP approach, we effectively "repair" all truth table entries that do not match the idempotence requirement - arbitrarily based on the first encountered entry in the truth table - thus making it easy to find such a function in any problem size.
However, that way, the link between the symbolic form of the function and the truth table that is evaluated is broken, since the truth table that is directly obtained from the GP expression is modified in this process.
This makes the GP ``blind'' to a large part of the search space, since many different Boolean expressions will be transformed into the same truth table.
The bitstring representation, on the other hand, operates directly in the restricted search space, greatly improving the performance.

Regarding the fitness functions, the advantage of additional information from the Walsh-Hadamard spectrum values, present in the second objective $fit_2$, is clearly visible in the results.
While in smaller function sizes the difference is not that pronounced, it becomes very useful to apply the second fitness function in a larger number of variables.
Unfortunately, the local search has not proven helpful, regardless of the representation. 
This is probably due to the implementation, which relies on available mutation operators, instead of exploring a specific neighborhood that might be defined in this context.

Dimension 9 represents the first dimension where we do not reach the best-known nonlinearity value. Nonlinearity 241 was achieved by Kavut et al., and the authors used specialized heuristics and searched in the rotation symmetric Boolean function class~\cite{cryptoeprint:2006/181}. Since we said that there is equivalence between idempotent and rotation symmetric functions when using the normal basis, this is a clear sign that the problem becomes more difficult in the polynomial basis, making our approach fail to reach 241. Naturally, since the operation of cyclic rotation (in case of rotation symmetric functions) is simpler than the linear map we need to consider for idempotent functions, this result is also not surprising. Nonlinearity 242 is achieved by considering the generalized rotation symmetric class~\cite{KAVUT2010341}, making it clear why we struggle to reach that value, as we obviously struggle to obtain rotation symmetric functions.
Larger sizes ($n>9$) also do not yield optimal results, which is somewhat surprising, at least for $n$ even, where we would expect bent functions to be easily found.
To conclude, restricting the search to idempotent functions makes the evolutionary search difficult. For the unrestricted encoding, the reason is clear: the disruptive nature of crossover and mutation operators makes it difficult to even find idempotent functions. However, restricted encoding enforces idempotence while making the search space much smaller. We hypothesize that the difficulty then comes from the generality of the allowed linear map, making too many solutions viable (i.e., idempotent) while not having enough pressure toward high nonlinearity.

\begin{table*}
\centering
\caption{Best obtained fitness values per representation/encoding}
\label{tab:bestnl}
\begin{tabular}{@{}lrrrrrrr@{}}
\toprule
n                & \multicolumn{1}{l}{6} & \multicolumn{1}{l}{7} & \multicolumn{1}{l}{8} & \multicolumn{1}{l}{9} & \multicolumn{1}{l}{10} & \multicolumn{1}{l}{11} & \multicolumn{1}{l}{12} \\ \midrule
GP, $fit1$        & 28                    & 16                    & 64                    & 2                     & 4                      & 2                      & 1024                   \\
GP, $fit2$        & 24                    & 16                    & 64                    & 2                     & 4                      & 2                      & 4                      \\
GP\_R, $fit1$     & 28                    & 56                    & 120                   & 240                   & 488                    & 984                    & 1983                   \\
GP\_R, $fit2$     & 28                    & 56                    & 120                   & 240                   & 488                    & 986                    & 1988                   \\
GP\_R, LS, $fit1$ & 28                    & 56                    & 120                   & 240                   & 487                    & 984                    & 1984                   \\
GP\_R, LS, $fit2$ & 28                    & 56                    & 120                   & 240                   & 486                    & 986                    & 1988                   \\
TT, $fit1$        & 28                    & 56                    & 112                   & -14                   & -64                    & -184                   & -474                   \\
TT, $fit2$        & 28                    & 56                    & 114                   & -37                   & -105                   & -271                   & -595                   \\
TT\_R, $fit1$     & 28                    & 56                    & 120                   & 240                   & 488                    & 986                    & 1992                   \\
TT\_R, $fit2$     & 28                    & 56                    & 120                   & 240                   & 488                    & 988                    & 1992                   \\
TT\_R, LS, $fit1$ & 28                    & 56                    & 120                   & 240                   & 488                    & 986                    & 1991                   \\
TT\_R, LS, $fit2$ & 28                    & 56                    & 120                   & 240                   & 488                    & 987                    & 1992                   \\ \bottomrule
\textit{Best-known}      & {28} & {56} & {120} & {242}  & {496} & {996}   & {2016}
\end{tabular}
\end{table*}

\section{Conclusions and Future Work}
\label{sec:conclusions}
This paper considered the design of idempotent Boolean functions with high nonlinearity via evolutionary algorithms. To this end, we proposed two different types of solution encodings, based on the truth table representation of Boolean functions (suitable for GA) and a syntactic tree representation (suitable for GP). Further, we designed a restricted encoding that forces the candidate solutions to be idempotent by letting the EA tweak only the representatives of the squaring orbits that define the idempotence property. The GA and GP were tested under two different fitness functions on the sets of Boolean functions of $n=5$ up to $n=12$ variables.

Despite the link with rotation symmetry (which has already been investigated in the literature, obtaining good performance with EA), our results showed that evolving highly nonlinear idempotent functions is considerably more difficult than evolving rotation symmetric functions. While in the unrestricted encoding case the reason is quite evident---the EA struggles in finding idempotent functions in the first place, let alone highly nonlinear ones---we observed that the evolutionary search becomes more difficult also in the restricted case for $n>9$ variables, which is more surprising.

The results obtained in this paper prompt several interesting open questions for future research. As a first general direction, it would be interesting to investigate more in depth the relationship between evolving idempotent and rotation-symmetric functions, considering the results obtained in prior literature. For example, in this work, we have observed that the best nonlinearity obtained by GA and GP for $n=9$ variables is 240, while other works~\cite{Kavut07,Carlet2025} achieved a nonlinearity of 241.

In particular, the result from~\cite{Carlet2025} was achieved on the rotation symmetric class by using a GA with restricted truth table encoding, enhanced with a local search step. On the other hand, our results showed that LS does not introduce any visible advantage in evolving highly nonlinear idempotent functions, independently of the representation. Since we used the same local search strategy adopted in~\cite{Carlet2025}, it could be interesting to experiment with other approaches for the local search step. For example, instead of using a fixed number of random mutations, one could apply a complete search of the neighborhood of each solution, in a hill-climbing fashion.

Designing a deterministic local search step could also be useful to sample the space of candidate solutions, and build the corresponding Local Optima Networks (LONs)~\cite{OchoaTVD08}. As remarked by the authors of~\cite{Carlet2025}, there are very few works addressing the fitness landscape analysis of Boolean functions and their cryptographic properties~\cite{JakobovicPMW21}. An interesting direction here would be to compare the LONs of rotation symmetric and idempotent functions, to check whether any useful insights on the performance difference of EA on these two optimization problems can be gained.

Finally, another interesting direction to explore concerns the investigation of further constraints on the set of idempotent functions. For instance, a concrete example would be to restrict the set of candidate solutions to balanced idempotent functions. This would require adapting the restricted encodings, or alternatively, coming up with additional penalty factors in the fitness functions. In the former case, one could then integrate or adapt specific crossover and mutation operators for generic balanced functions (like those proposed in~\cite{manzoni20}) to the setting of idempotent functions.

\bibliographystyle{abbrv}
\bibliography{bibliography}

\begin{thebibliography}{10}

\bibitem{burnside}
W.~Burnside.
\newblock Theory of groups of finite order.
\newblock {\em Messenger of Mathematics}, 23:112, 1909.

\bibitem{carlet_2021}
C.~Carlet.
\newblock {\em Boolean Functions for Cryptography and Coding Theory}.
\newblock Cambridge University Press, Cambridge, 2021.

\bibitem{CarletDGJMP24}
C.~Carlet, M.~Durasevic, B.~Gasperov, D.~Jakobovic, L.~Mariot, and S.~Picek.
\newblock A new angle: On evolving rotation symmetric boolean functions.
\newblock In S.~L. Smith, J.~Correia, and C.~Cintrano, editors, {\em
  Applications of Evolutionary Computation - 27th European Conference,
  EvoApplications 2024, Held as Part of EvoStar 2024, Aberystwyth, UK, April
  3-5, 2024, Proceedings, Part {I}}, volume 14634 of {\em Lecture Notes in
  Computer Science}, pages 287--302. Springer, 2024.

\bibitem{10.1007/978-3-031-70085-9_27}
C.~Carlet, M.~Durasevic, D.~Jakobovic, and S.~Picek.
\newblock Discovering rotation symmetric self-dual bent functions with
  evolutionary algorithms.
\newblock In M.~Affenzeller, S.~M. Winkler, A.~V. Kononova, H.~Trautmann,
  T.~Tu{\v{s}}ar, P.~Machado, and T.~B{\"a}ck, editors, {\em Parallel Problem
  Solving from Nature -- PPSN XVIII}, pages 429--445, Cham, 2024. Springer
  Nature Switzerland.

\bibitem{Carlet2025}
C.~Carlet, M.~Durasevic, D.~Jakobovic, S.~Picek, and L.~Mariot.
\newblock A systematic study on the design of odd-sized highly nonlinear
  boolean functions via evolutionary algorithms.
\newblock {\em Genetic Programming and Evolvable Machines}, 27(1), Dec. 2025.

\bibitem{Djurasevic2023}
M.~Djurasevic, D.~Jakobovic, L.~Mariot, and S.~Picek.
\newblock A survey of metaheuristic algorithms for the design of cryptographic
  {B}oolean functions.
\newblock {\em Cryptography and Communications}, 15(6):1171--1197, July 2023.

\bibitem{FullerDM03}
J.~Fuller, E.~Dawson, and W.~Millan.
\newblock Evolutionary generation of bent functions for cryptography.
\newblock In {\em Proceedings of the {IEEE} Congress on Evolutionary
  Computation, {CEC} 2003, Canberra, Australia, December 8-12, 2003}, pages
  1655--1661. {IEEE}, 2003.

\bibitem{10.1007/978-3-319-10762-2_41}
R.~Hrbacek and V.~Dvorak.
\newblock Bent function synthesis by means of cartesian genetic programming.
\newblock In T.~Bartz-Beielstein, J.~Branke, B.~Filipi{\v{c}}, and J.~Smith,
  editors, {\em Parallel Problem Solving from Nature -- PPSN XIII}, pages
  414--423, Cham, 2014. Springer International Publishing.

\bibitem{10.1145/3067695.3084220}
J.~Husa and R.~Dobai.
\newblock Designing bent {B}oolean functions with parallelized linear genetic
  programming.
\newblock In {\em Proceedings of the Genetic and Evolutionary Computation
  Conference Companion}, GECCO '17, page 1825–1832, New York, NY, USA, 2017.
  Association for Computing Machinery.

\bibitem{JakobovicPMW21}
D.~Jakobovic, S.~Picek, M.~S.~R. Martins, and M.~Wagner.
\newblock Toward more efficient heuristic construction of boolean functions.
\newblock {\em Appl. Soft Comput.}, 107:107327, 2021.

\bibitem{Kavut07}
S.~Kavut, S.~Maitra, and M.~D. Y{\"{u}}cel.
\newblock Search for boolean functions with excellent profiles in the rotation
  symmetric class.
\newblock {\em {IEEE} Trans. Inf. Theory}, 53(5):1743--1751, 2007.

\bibitem{cryptoeprint:2006/181}
S.~Kavut, S.~Maitra, and M.~D. Yücel.
\newblock There exist boolean functions on $n$ (odd) variables having
  nonlinearity $> 2^{n-1} - 2^{\frac{n-1}{2}}$ if and only if $n > 7$.
\newblock Cryptology ePrint Archive, Paper 2006/181, 2006.
\newblock \url{https://eprint.iacr.org/2006/181}.

\bibitem{KAVUT2010341}
S.~Kavut and M.~D. Yücel.
\newblock 9-variable boolean functions with nonlinearity 242 in the generalized
  rotation symmetric class.
\newblock {\em Information and Computation}, 208(4):341--350, 2010.

\bibitem{KERDOCK1972182}
A.~Kerdock.
\newblock A class of low-rate nonlinear binary codes.
\newblock {\em Information and Control}, 20(2):182 -- 187, 1972.

\bibitem{4729749}
W.~M. Liu and A.~Youssef.
\newblock On the existence of $(10, 2, 7, 488)$ resilient functions.
\newblock {\em IEEE Transactions on Information Theory}, 55(1):411--412, 2009.

\bibitem{MacWilliams-Sloane}
F.~J. MacWilliams and N.~J.~A. Sloane.
\newblock {\em The Theory of Error-Correcting Codes}.
\newblock Elsevier, Amsterdam, North Holland, 1977.
\newblock {ISBN: 978-0-444-85193-2}.

\bibitem{manzoni20}
L.~Manzoni, L.~Mariot, and E.~Tuba.
\newblock Balanced crossover operators in genetic algorithms.
\newblock {\em Swarm Evol. Comput.}, 54:100646, 2020.

\bibitem{MariotSLM22}
L.~Mariot, M.~Saletta, A.~Leporati, and L.~Manzoni.
\newblock Heuristic search of (semi-)bent functions based on cellular automata.
\newblock {\em Nat. Comput.}, 21(3):377--391, 2022.

\bibitem{OchoaTVD08}
G.~Ochoa, M.~Tomassini, S.~V{\'{e}}rel, and C.~Darabos.
\newblock A study of {NK} landscapes' basins and local optima networks.
\newblock In C.~Ryan and M.~Keijzer, editors, {\em Genetic and Evolutionary
  Computation Conference, {GECCO} 2008, Proceedings, Atlanta, GA, USA, July
  12-16, 2008}, pages 555--562. {ACM}, 2008.

\bibitem{1056589}
J.~{Olsen}, R.~{Scholtz}, and L.~{Welch}.
\newblock Bent-function sequences.
\newblock {\em IEEE Transactions on Information Theory}, 28(6):858--864,
  November 1982.

\bibitem{00190}
S.~Picek, C.~Carlet, S.~Guilley, J.~F. Miller, and D.~Jakobovic.
\newblock Evolutionary algorithms for boolean functions in diverse domains of
  cryptography.
\newblock {\em Evolutionary Computation}, 24(4):667--694, 2016.

\bibitem{10.1145/2908812.2908915}
S.~Picek and D.~Jakobovic.
\newblock Evolving algebraic constructions for designing bent {B}oolean
  functions.
\newblock In {\em Proceedings of the Genetic and Evolutionary Computation
  Conference 2016}, GECCO ’16, page 781–788, New York, NY, USA, 2016.
  Association for Computing Machinery.

\bibitem{10.1007/978-3-319-16501-1_16}
S.~Picek, D.~Jakobovic, J.~F. Miller, E.~Marchiori, and L.~Batina.
\newblock Evolutionary methods for the construction of cryptographic {B}oolean
  functions.
\newblock In P.~Machado, M.~I. Heywood, J.~McDermott, M.~Castelli,
  P.~Garc{\'i}a-S{\'a}nchez, P.~Burelli, S.~Risi, and K.~Sim, editors, {\em
  Genetic Programming}, pages 192--204, Cham, 2015. Springer International
  Publishing.

\bibitem{poli08:fieldguide}
R.~Poli, W.~B. Langdon, and N.~F. McPhee.
\newblock {\em A Field Guide to Genetic Programming}.
\newblock lulu.com, 2008.

\bibitem{Rothaus}
O.~Rothaus.
\newblock On “bent” functions.
\newblock {\em Journal of Combinatorial Theory, Series A}, 20(3):300 -- 305,
  1976.

\bibitem{a16110499}
L.~Rovito, A.~De~Lorenzo, and L.~Manzoni.
\newblock Discovering non-linear boolean functions by evolving walsh transforms
  with genetic programming.
\newblock {\em Algorithms}, 16(11), 2023.

\bibitem{StanicaMC04}
P.~Stănică, S.~Maitra, and J.~A. Clark.
\newblock Results on rotation symmetric bent and correlation immune boolean
  functions.
\newblock In B.~K. Roy and W.~Meier, editors, {\em Fast Software Encryption,
  11th International Workshop, {FSE} 2004, Delhi, India, February 5-7, 2004,
  Revised Papers}, volume 3017 of {\em Lecture Notes in Computer Science},
  pages 161--177, Berlin, Heidelberg, 2004. Springer.

\bibitem{Wang2022}
Y.~Wang, G.~Gao, and Q.~Yuan.
\newblock Searching for cryptographically significant rotation symmetric
  boolean functions by designing heuristic algorithms.
\newblock {\em Security and Communication Networks}, 2022:1--6, Mar. 2022.

\bibitem{10.1145/3579856.3590337}
L.~Yan, J.~Cui, J.~Liu, G.~Xu, L.~Han, A.~Jolfaei, and X.~Zheng.
\newblock Iga: An improved genetic algorithm to construct weightwise (almost)
  perfectly balanced boolean functions with high weightwise nonlinearity.
\newblock In {\em Proceedings of the 2023 ACM Asia Conference on Computer and
  Communications Security}, ASIA CCS '23, page 638–648, New York, NY, USA,
  2023. Association for Computing Machinery.

\end{thebibliography}

\end{document}